%% file: main_journal.tex
\documentclass[10pt,journal,compsoc, dvipsnames]{IEEEtran}
\usepackage[square,sort,comma,numbers]{natbib}

\usepackage{natbib}
\usepackage{color}
\usepackage{wrapfig}
\usepackage{booktabs} 
\usepackage{amsmath,amssymb}
\usepackage{tikz}
\usepackage{bbm}
\usepackage{subcaption, graphicx}
\usepackage{multirow}

\usepackage{standalone}

\newcommand{\norm}[1]{\left\lVert#1\right\rVert}

\newcommand{\kld}[2]{\ensuremath{KL\left[{#1} \parallel {#2}\right]}}

\DeclareMathOperator{\E}{\mathbb{E}}

\ifCLASSINFOpdf
\else
\fi
\begin{document}
%
% paper title
% Titles are generally capitalized except for words such as a, an, and, as,
% at, but, by, for, in, nor, of, on, or, the, to and up, which are usually
% not capitalized unless they are the first or last word of the title.
% Linebreaks \\ can be used within to get better formatting as desired.
% Do not put math or special symbols in the title.
\title{Learned Image Compression for \\ \textit{Machine} Perception}
%
%
% author names and IEEE memberships
% note positions of commas and nonbreaking spaces ( ~ ) LaTeX will not break
% a structure at a ~ so this keeps an author's name from being broken across
% two lines.
% use \thanks{} to gain access to the first footnote area
% a separate \thanks must be used for each paragraph as LaTeX2e's \thanks
% was not built to handle multiple paragraphs
%
%
%\IEEEcompsocitemizethanks is a special \thanks that produces the bulleted
% lists the Computer Society journals use for "first footnote" author
% affiliations. Use \IEEEcompsocthanksitem which works much like \item
% for each affiliation group. When not in compsoc mode,
% \IEEEcompsocitemizethanks becomes like \thanks and
% \IEEEcompsocthanksitem becomes a line break with idention. This
% facilitates dual compilation, although admittedly the differences in the
% desired content of \author between the different types of papers makes a
% one-size-fits-all approach a daunting prospect. For instance, compsoc 
% journal papers have the author affiliations above the "Manuscript
% received ..."  text while in non-compsoc journals this is reversed. Sigh.

\author{Felipe Codevilla $^{2, 4}$,
        Jean Gabriel Simard $^{2}$,
        Ross Goroshin$^{5}$ and
        Chris Pal$^{1, 2, 3, 6}$% <-this % stops a space
          %\And 
          \vspace{4pt}\\
  \footnotesize{$^{1}$Polytechnique Montréal,
                $^{2}$Mila, Quebec AI Institute
                $^{3}$ElementAI / Service Now,} \\
  \footnotesize{$^{4}$Independent Robotics, 
                $^{5}$Google Brain Inc.,} \\
  \footnotesize{$^{6}$Canada CIFAR AI Chair, \\
            Correspondence: felipe.alcm@gmail.com} 
  %\footnotesize{\textbf{\url{https://fgolemo.github.io/autobots/}}
  }

\IEEEtitleabstractindextext{%
\begin{abstract}
Recent work has shown that learned image compression strategies can outperform standard hand-crafted compression algorithms that have been developed over decades of intensive research on the rate-distortion trade-off. With growing  applications of computer vision, high quality image reconstruction from a compressible representation is often a secondary objective. Compression that ensures high accuracy on computer vision tasks such as image segmentation, classification, and detection therefore has the potential for significant impact across a wide variety of settings. In this work, we develop a framework that produces a compression format suitable for both human perception and machine perception. We show that representations can be learned that simultaneously optimize for compression and performance on core vision tasks. Our approach allows models to be trained directly from compressed representations, and this approach yields increased performance on new tasks and in low-shot learning settings.  We present results that improve upon segmentation and detection performance compared to standard high quality JPGs, but with representations that are \textbf{four to ten times smaller} in terms of bits per pixel. Further, unlike naive compression methods, at a level ten times smaller than standard JEPGs, segmentation and detection models trained from our format suffer only minor degradation in performance. 
\end{abstract}
% Note that keywords are not normally used for peerreview papers.
\begin{IEEEkeywords}
Image Compression, Computer Vision
\end{IEEEkeywords}}
% make the title area
\maketitle
% To allow for easy dual compilation without having to reenter the
% abstract/keywords data, the \IEEEtitleabstractindextext text will
% not be used in maketitle, but will appear (i.e., to be "transported")
% here as \IEEEdisplaynontitleabstractindextext when the compsoc 
% or transmag modes are not selected <OR> if conference mode is selected 
% - because all conference papers position the abstract like regular
% papers do.
%\IEEEdisplaynontitleabstractindextext
% \IEEEdisplaynontitleabstractindextext has no effect when using
% compsoc or transmag under a non-conference mode.
% For peer review papers, you can put extra information on the cover
% page as needed:
% \ifCLASSOPTIONpeerreview
% \begin{center} \bfseries EDICS Category: 3-BBND \end{center}
% \fi
%
% For peerreview papers, this IEEEtran command inserts a page break and
% creates the second title. It will be ignored for other modes.
%\IEEEpeerreviewmaketitle
\IEEEraisesectionheading{\section{Introduction}\label{sec:introduction}}
% The very first letter is a 2 line initial drop letter followed
% by the rest of the first word in caps (small caps for compsoc).
% 
% form to use if the first word consists of a single letter:
% \IEEEPARstart{A}{demo} file is ....
% 
% form to use if you need the single drop letter followed by
% normal text (unknown if ever used by the IEEE):
% \IEEEPARstart{A}{}demo file is ....
% 
% Some journals put the first two words in caps:
% \IEEEPARstart{T}{his demo} file is ....
% 
% Here we have the typical use of a "T" for an initial drop letter
% and "HIS" in caps to complete the first word.
\IEEEPARstart{I}{t} has been shown that image classification can be performed directly from the discrete cosine transformations used in the standard JPEG compression format, that this procedure is more computationally efficient and maintains performance comparable to state-of-the-art methods for image classification tasks \cite{gueguen2018faster}. In recent years a growing research community has emerged around learned image compression strategies based on deep neural networks, and many results have demonstrated that such approaches can surpass highly engineered approaches to image compression \cite{Agustsson2018, Balle2018, minnen2018joint}. 

%In contrast to standard compression methods, learned compressors can be more easily adapted to specific tasks and domains such as medical image analysis and autonomous driving.

 \begin{figure*}[!ht]
       \centering
         \begin{tabular}{cc}
            \includegraphics[width=0.45\linewidth]{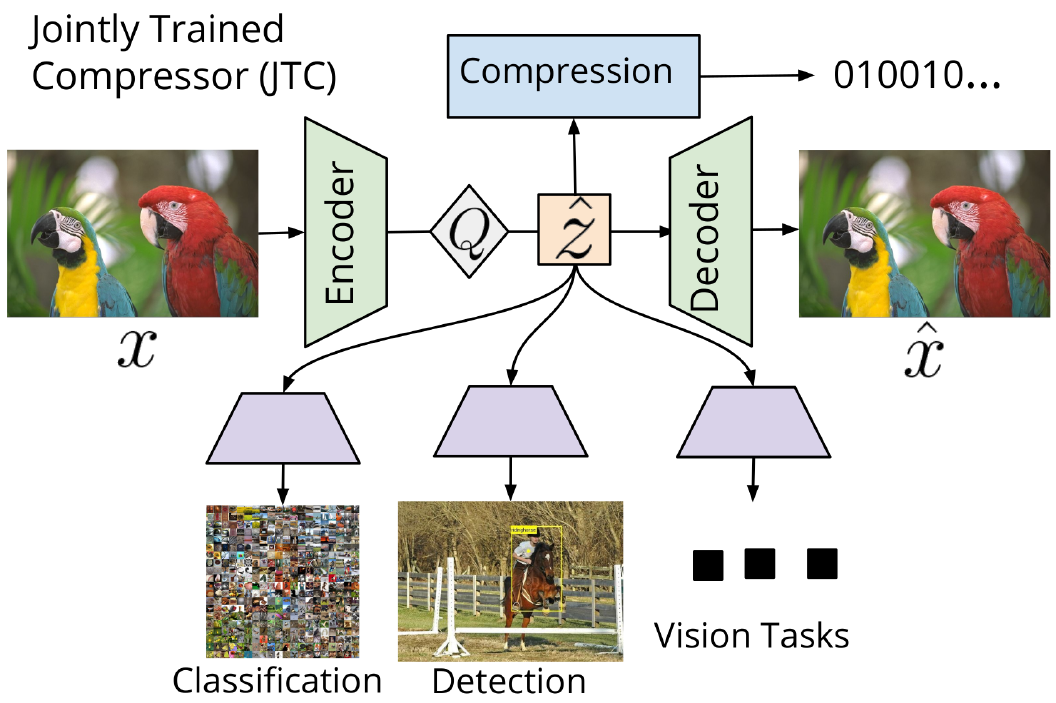} &
          \includegraphics[width=0.50\linewidth]{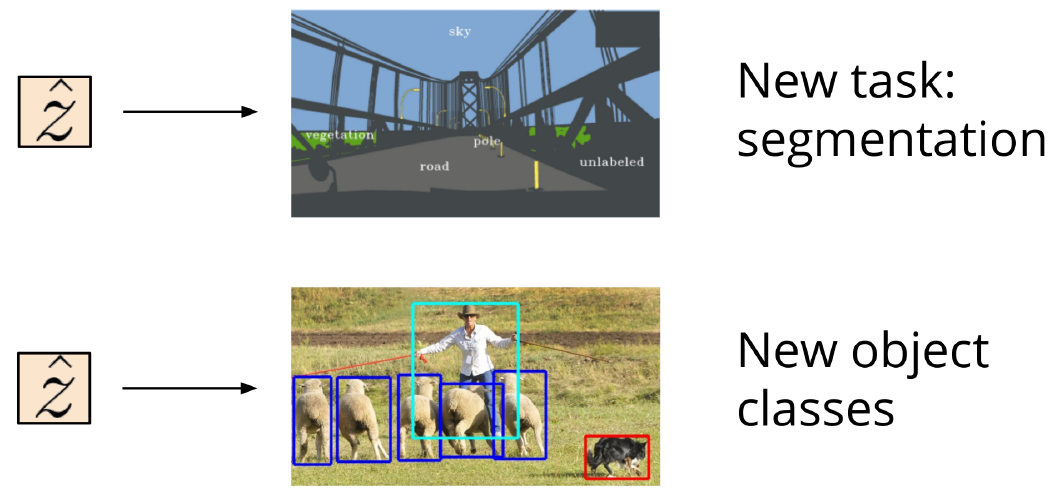}
          \\ %\includegraphics[width=0.39\linewidth]{figures/architechture/Downstreamtasks.png} \\
            \footnotesize{(a) We jointly train an image  compressor.} &  \footnotesize{(b) \textbf{New} tasks are trained from \textbf{fixed} compressed format.}
          \end{tabular}
               
       \caption{A high level view of our strategy for learning compressible image representations for machine perception. We jointly train a model to compress as well as perform key vision tasks directly from the compressible representation. The approach leads to significant gains compared to both JPEG or naive learned deep compression when training new models for new tasks directly from our learned compressible representations. }
   \label{fig:general_view}
 \end{figure*}

Neural image compression is normally formulated as an optimization balancing a trade-off between two objectives: the compression ratio and image distortion (or image quality). A common way to achieve this is to produce a representation that contains the information necessary for reconstruction while maintaining low entropy on the latent representation, under some probability model. This latent space can then be compressed to a bit-stream using a lossless compression algorithm, thus achieving a lower BPP compression of the image. In contrast, in our work here we study whether this latent space representation can be further biased or re-structured such that it better represents and retains information that is relevant for subsequent semantically oriented vision tasks such as classification, detection, and segmentation. 
Intuitively, our work aims to develop an approach to learning neural image compression in a way that preserves or improves performance on common machine perception tasks and yields representations that likely to retain information required when training models for \emph{new} tasks - since the representation is explicitly learned such that it is capable of reconstructing the original image. In our framework, these different tasks are performed by task specific decoders that operate directly on the (common) learned, compressible representation. Furthermore, we test the hypothesis that our neural compression models learned in this way might be particularly good at adapting to new tasks using a competitive low-show learning benchmark evaluation. Our approach yields state of the art performance on this evaluation. A high level view of our approach is shown in Figure \ref{fig:general_view}.

We explore these ideas by biasing the learned representations of neural encoders specifically constructed to yield quantizable and compressible codes (under arithmetic coding) by jointly training them with alternative decoders for auxiliary tasks. We conjecture that this should lead to compressible representations having comparable rate/distortion characteristics while better preserving and structuring the information necessary for performing computer vision tasks directly from these highly compressible codes. We underscore that this is potentially a challenging setting, since other transfer learning work has shown that computer vision tasks such as auto-encoding and semantic segmentation can be less compatible than others, leading to decreased performance when models are trained on both tasks simultaneously \cite{zamir2018taskonomy}. 

%Compressible formats have two key advantages: 1) they may be lower dimensional than the input images, leading to computational and memory savings during inference in computer vision applications, and 2) they can be compressed to even lower BPP (bits-per-pixel) under subsequent lossless compression algorithms. 

The contribution of our work here is to propose and explore a joint training procedure that allows models to learn highly compressible representations, while minimizing the impact on performance on key machine perception tasks. Importantly, in our approach these tasks are performed directly from the compressed representation, without the need for decoding the image.  We test our hypothesis that this approach might yield representations that are well suited for new tasks and few-shot learning. We find that they achieve state-of-the-art performance on few-shot image detection when learning directly from our task-informed compressible representations.  Furthermore, our representations can be compressed to be \emph{four to ten times smaller than standard JPEG imagery} while improving performance or suffering only a minor degradation on the key vision tasks of classification, detection and segmentation.  %Our PyTorch code is available at \url{https://github.com/anonymouscodebase}.

% This needs to be cleaned and we need the paragraph which says, our contributions here are:

%The trade-off between only compression and distortion is a limiting assumption. The distortion is usually limited by visual perception. \red{Key components of the function are semantically meaningful which is different than perception, we may optimize for the semantics that exists on a specific task.}

% Not sure if we need this
% \begin{figure}[!th]
%       \includegraphics[width=0.8\linewidth]{figures/Downstreamtasks.png}
%       \caption{Training downstream tasks from the compressed representation.}
%   \label{fig:downstream_general_view}
% \end{figure}

%\input{figures/architechture/architecture.tex}

\section{Learning Compressible Representations for Machine Perception Tasks}
% In this section to develop the theory of a semantically biased compressed representations. This theory leads us to introduce a network architecture for learning compressed representations biased towards machine vision tasks. In this section we posit that there is a multitude of compression formats, some that are more amenable to semantic tasks. 
\subsection{Preliminaries: Rate-Distortion and Perceptual Quality}
%\subsection{Image Compression for Machine Perception}
The traditional goal of lossy image compression is to find the best trade-off between minimizing the expected length of the bit-stream representing the input image, $x$, and minimizing the expected distortion between the image and its reconstruction $\hat{x}$. 
In the information theory literature, this is known as the rate-distortion optimization problem \cite{cover1999elements}.
In the context of learned image compression, this can be formulated as a multi-task loss: 
\begin{equation}%\label{eq:rd}
    \mathcal{R}(\mathbf{Q}(f(x))) + \lambda \cdot \mathcal{D}(g(\mathbf{Q}(f(x))), x),
    \label{eq:r_and_d}
\end{equation}
where $\mathcal{R}$ represents the rate or 
%the encoded information function, proportional to 
bit-stream length, and $\mathcal{D}$ represents the distortion or reconstruction fidelity, usually implemented by a distance function. To find such an encoding, the non-linear transform coding strategy \cite{goyal2001theoretical} consists of producing a discrete latent representation $z$ from an image $x$ using an \textit{encoder}, i.e. $z = f(x)$. This real-valued representation is quantized, by a quantization operator $\mathbf{Q}$, into $\hat{z}$. 
The reconstructed input, $\hat{x}$, is generated by the image decoder $g(\hat{z})$. Because both objectives cannot be completely minimized simultaneously, different values of $\lambda$ lead to a different operating points in the trade-off between rate and distortion. Note that the representation $\hat{z}$ must be an element of a finite set to be losslessly encoded to a bit-stream with a lossless compression algorithm such as arithmetic coding \cite{langdon1984introduction}. 
%
% perception part
Because perceptual quality is not directly correlated with distortion, a perceptual quality metric is often added to the loss \citep{blau2019rethinking}.  This idea has been used to improve the quality of image compression algorithms \citep{mentzer2020high}. 
% Formally, perceptual quality can be defined as the divergence between the reconstructed image distribution $p_{\hat{x}}$, and the natural image signal distribution, $d(p_x,p_{\hat{x}})$. 

% Something wrong written on the towards paer is that they actually use the encoder 

\subsection{Rate-Distortion-Utility Perspective}

%\subsection{Compressed Inference}
% Introducing compressed inference

The representation $z$ can be encoded in a bit stream
and stored or transmitted, but also used directly to perform visual tasks \textit{without reconstructing} the input image \citep{torfason2018towards}. This common use-case highlights the potential of exploring another dimension beyond the classical rate-distortion optimization trade-off. Rather than thinking of perceptual quality as the difference between reconstructed and the original images, we are interested in the direct utility of the learned compressed representation for machine perception tasks. 

%Adversarial attacks do not change the general appearance of images but can often drastically alter machine perception performance. For example it was observed that imperceptible changes to the image can have a substantial impact on image classification performance \citep{dodge2016understanding}. The mentioned evidence conveys the idea that vision task performance is not necessarily proportional to perceptual quality and distortion. 

This leads to our \emph{key hypothesis}: %that quantifies the how semantically usable is the encoded information.
% Maybe this is a claim.
%
%\textbf{Hypothesis:}
For any desired rate, $R$, and distortion level $D$, we conjecture that there exists many different, quantized latent representations, $\hat{z}=\mathbf{Q}(f(x))$, which lead to significant differences in performance on subsequent machine vision tasks. In other words, there exists many representations which achieve a given rate-distortion trade-off, but some representations may be better suited for subsequent semantic tasks than others.
%\label{def:main}
%\end{theorem}

%An analogous view, in order to better understand Claim \ref{def:main} is
%In light of our hypothesis, we view our setting as a constrained representation learning problem 
%\begin{equation}
%    h(f(x)) , \\ \; s.t \,\; r(\mathbf{Q}(f(x)),\hat{x}) < R,  \;  d(g(\mathbf{Q}(f(x))),\hat{x}) < D,
%    \label{eq:constraint}
%\end{equation}
%where $f(x)$ is a representation we want to learn which respects the rate and distortion restrictions, $(R,D)$, and $h(\cdot)$ is a task-specific decoder that receives the learned representation as input. Constraint \ref{eq:constraint} guarantees that the representation produced by $f(x)$ is at least partially inevitable to recover the original image with a distortion error $D$, and, can be losslessly encoded into a bitstream with a rate $R$. %Those constraints improve a full cross task generalization since the representation learned is compact and holds all the spatial features necessary to reconstruct back the image at a certain distortion. This was already observed for auto-encoders \citep{le2018supervised} but here we add also the compression constraint.
In this context, we propose that the objective function to be optimized should go beyond the classical rate-distortion paradigm, to include a measure of the utility of the representation for concrete machine perception tasks:
\begin{equation}
    \mathcal{R}(\mathbf{Q}(f(x)), x) + \lambda_d \cdot \mathcal{D}(g(\mathbf{Q}(f(x))), x) + \lambda_u \cdot \mathcal{U}(\mathbf{Q}(f(x))), \label{eq:RDU}
\end{equation}
where  $\mathcal{U}(\cdot)$ is a pragmatically defined utility, or machine perceptual quality metric.
In contrast to the usual notion of human perceptual quality which is often characterized by metrics such as squared error, we can define machine vision perceptual quality much more pragmatically, i.e. based on a multi-task loss:
\begin{equation}
    \mathcal{U}(\mathbf{Q}(f(x))) =
    \sum_{t\in\mathcal{T}} \lambda_t [L_{t}(h_t(f(x)), y_t)], \label{eq:u_tasks}
\end{equation}
where $L_{t}$ is a loss function for an specific task $t$ with respect to the labels
$y_t$. The functions $h_t$ are the task specific decoders. We conjecture that the optimization of learned compression models for rate $\mathcal{R}$, distortion $\mathcal{D}$, and utility $\mathcal{U}$ as defined in (\ref{eq:RDU}) will result in encoders $f(x)$ that are capable of producing low bit-per-pixel quantized representations $\hat{z}$ that are better structured to facilitate the learning of new machine perception tasks directly from $\hat{z}$. 
% Todo extend on why informed z might be better than pure z

\subsection{Compression Architecture}

%Our approach to implement problem uses non-linear transform coding strategy \cite{goyal2001theoretical} which consists in producing a discrete latent representation $\hat{z}$ from an an image $x$ using an \textit{encoder} $f_{\theta_e}$, followed by lossless compression (e.g. arithmetic coding) of the latent representation to a bit-stream
% We need to define the mechanism to compute the encoded information function $r$, defined on Eq. \ref{eq:r_and_d}. For this we use the probabilistic model $p_{\hat{z}}$. The optimization problem corresponding to this approach can be formulated as:
% \begin{equation}\label{eq:rd}
%     \min R + \lambda D = \E[-\log_{2}p_{\hat{z}}(\hat{z})] + \lambda \E[d(x, \hat{x})],
% \end{equation}
% where the first expectation represents the expected length of the compressed bit-stream and the second expectation is the expected distortion between the input and its reconstruction by the image decoder $g_{\theta_i}$.% A different value of $\lambda$ leads to a different operation point in the trade-off between rate and distortion. The representation $\hat{z}$ must be an element of a finite set to be losslessly encoded to a bit-stream using a lossless compression algorithm such as arithmetic coding \cite{langdon1984introduction}. 
We base our image compression approach on the hierarchical variational-auto-encoder (VAE) formulation proposed by \cite{balle2018variational}. The function $\mathcal{R}$ is computed as the expected marginal probability of the latent \emph{quantized} representation, $\E[-\log_{2}p_{\hat{z}}(\hat{z})]$, where $\hat{z}$ is a the quantized encoding of an input image and the expectation is with respect to the empirical distribution of a training set.
%The approach consists of a VAE compression framework that is forced to extract a low entropy latent representation ($z$) of the input ($x$), which is then quantized ($\hat{z})$ and losslessly compressed using arithmetic coding. 
A Gaussian Mean-Scale Model (GMSM) is used for $p_{\hat{z}}(\hat{z})$ as suggested in \citep{balle2018variational,minnen2018joint}. Each element of $\hat{z}$ is modeled by a Gaussian with a mean and a unique scale for a given lossless compression ratio.

%For further details about the compression framework, see \citep{balle2017}.
Since quantization is not a differentiable operation, a special form of uniform noise is added to the latent variables $z$ during training, which yields a differentiable strategy in which one must ensure that the amplitude of the noise keeps the codes within the same quantization bin. This induces the encoder to learn to output a code that is invariant to integer quantization. Recall that the addition of two independent random variables follows the convolution of their individual distributions. Therefore, the addition of uniform noise to the latent representation leads to a model of the form:
\begin{equation}
    p_{\hat{z}}(\hat{z}) = \prod_i \left(\mathcal{N}(\mu_i, \sigma_i ) *  u\left(-\frac{1}{2}, \frac{1}{2}\right)\right)(\hat{z}_i),
\end{equation}
where all the elements of $p_{\hat{z}}(\hat{z})$ are independent and `$*$' denotes the convolution operator.
The values of $\mu_i$ and $\sigma_i$ are the mean an variances output by the \textit{hyper-prior} network comprised of a separate VAE which encodes the latent representation $z$ into the quantized hyper-parameter $\hat{w}$ and decodes it into the parameters of the probabilisitic model $p_{\hat{z}}$. See Appendix I for further details.
%It can be shown that the probability distribution resulting from the latent distribution with additive noise reduces to the following difference of $F_{\hat{z}}$, the cumulative function of GMSM, around the quantized representation 
% \begin{equation}
%     p_{\hat{z}_i}(\hat{z}_i) =  F_{\mathcal{N}}\left(\hat{z}_i + \frac{1}{2} \middle| \mu_i, \sigma_i \right) - F_{\mathcal{N}}\left(\hat{z}_i - \frac{1}{2} \middle| \mu_i, \sigma_i \right)
% \end{equation}

We use a mean squared error distortion metric, which results in the following loss function for our naive (i.e. in that it is not task informed) compression framework:
\begin{equation}
    \label{eq:compression_optimization}
     %\Big[
     \E[-\log_{2}p_{\hat{z}}(\hat{z})] + \E[-\log_{2}p_{\hat{w }}(\hat{w})] + \lambda \E[\norm{x - g(\hat{z})}^2], %\Big],
\end{equation}
where $\E[-\log_{2}p_{\hat{z}}(\hat{z})]$ is the expected compression rate of the latent representation $\hat{z}$ and $\E[-\log_{2}p_{\hat{w}}(\hat{w})]$ the expected compression rate of the hyper-latent representation $\hat{w}$.
%Our approach to this problem uses non-linear transform coding strategy \cite{goyal2001theoretical} which consists in producing a discrete latent representation $\hat{z}$ from an an image $x$ using an \textit{encoder} $f_{\theta_e}$, followed by lossless compression (e.g. arithmetic coding) of the latent representation to a bit-stream using a probabilistic model $p_{\hat{z}}$. The optimization problem corresponding to this approach can be formulated as:
%\begin{equation}\label{eq:rd}
%    \min R + \lambda D = \E[-\log_{2}p_{\hat{z}}(\hat{z})] + \lambda \E[d(x, \hat{x})],
%\end{equation}
%where the first expectation represents the expected length of the compressed bit-stream and the second expectation is the expected distortion between the input and its reconstruction by the image decoder $g_{\theta_i}$. A different value of $\lambda$ leads to a different operation point in the trade-off between rate and distortion. %The representation $\hat{z}$ must be an element of a finite set to be losslessly encoded to a bit-stream using a lossless compression algorithm such as arithmetic coding \cite{langdon1984introduction}. 
The loss corresponding to our task informed compression setup is obtained
by combining the multi-task loss from (\ref{eq:u_tasks}) and using our rate-distortion-utility framework to obtain our task informed learned compression optimization framework:

\begin{multline}\label{eq:final_semantic_compression}
   %\Big[ 
   \E[-\log_{2}p_{\hat{z}}(\hat{z})] + \E[-\log_{2}p_{\hat{w}}(\hat{w} )] + 
  \lambda_e \E[\norm{x - g(\hat{x})}^2]  \\ +
    \sum_{t\in\mathcal{T}} \lambda_t \E[L_{t}(h_t(\hat{z}), y_t)]. 
    %\Big].
\end{multline}

\subsection{A Multi-task Inference and Compression Architecture}

 \begin{figure*}[!h]
     \centering
       \includegraphics[width=1.0\linewidth]{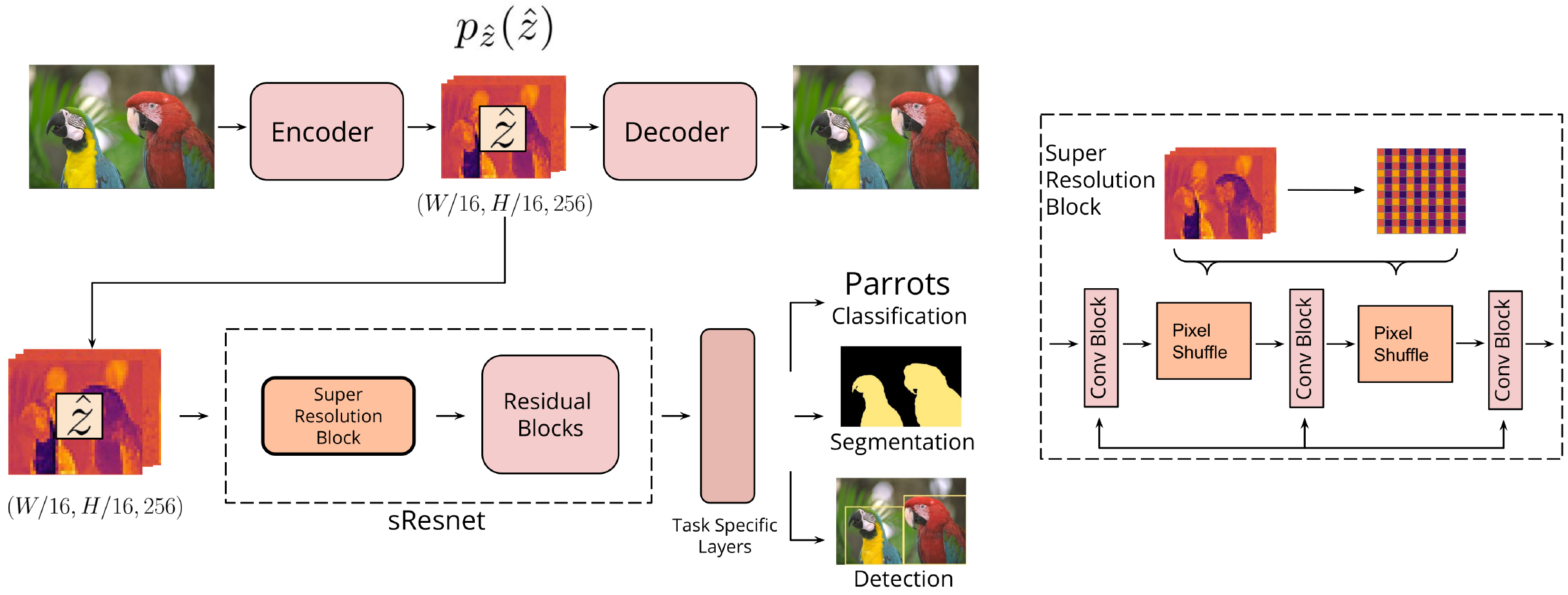}
   \caption {Detailed architecture of the proposed model for jointly training of image compression with different computer vision tasks.}
   \label{fig:architecture}
\end{figure*}
% TODO, we need to be clear about this two step strategy.

Here we illustrate the architecture, $h(z)$, used to produce results for different vision tasks. The same architecture is used both to optimize for the machine perception quality 
and perform inference from the compressible representation.

A detailed visualization of our proposed architecture is depicted in Figure \ref{fig:architecture}.  % show only the tasks specific part.
The compression architecture is depicted at the top of the figure (encoder/decoder). This architecture is the same as proposed by \cite{cheng2020learned}. The encoder is composed of four convolutional layers interlaced by so-called Generalized Divisive Normalization (GDN) layers \cite{Balle2016} for activation. The encoder expects an RGB image of size $3 \times W \times H$ and outputs a tensor of shape $256 \times W/16\times H/16$. Note, the rank of the compressible format is one third of the input image rank, but is actually smaller still (in bpp) because of it's redundancy. %The image decoder (left side, Fig. \ref{fig:architecture}) architecture is also equivalent to the one presented on \cite{balle2018variational}.
% The hyperprior architecture is identical as the one used by \cite{minnen2018joint}.
In order to perform vision tasks directly from the latent space, it is necessary to adapt the backbone ResNet architecture, commonly used to extract useful features for vision tasks. Compared to the original image, the input is comprised of many channels with lower spatial resolution.
Several solutions have been proposed to this problem. In \citep{gueguen2018faster}, the authors observed that a learned deconvolution yields better results since it helps overcoming the reduction in receptive field. In \cite{torfason2018towards}, the authors proposed to simply remove the the first layers of a Resnet and input the compressed space as is.
This inputs the latent space directly to the first Resnet bottleneck block,
which does not allow a space with bigger receptive field to be included.

Here we observed the necessity of recovering the full
image resolution and propose a pixel shuffle adapter that
outputs directly to the bottleneck blocks of the Resnet.
It was observed in \cite{torfason2018towards}, especially for lower bit rate configurations, a notable loss in performance for tasks that require spatial information such as semantic segmentation.
Pixel shuffle \cite{shi2016real} performs sub-pixel resolution 
convolutions and mitigates any potential loss of spatial information.
The architecture, refereed as \textit{sResnet} is depicted Fig. \ref{fig:architecture},
and contains three convolutional blocks (which include ReLUs) interlaced with two pixel shuffle blocks. A complete detailed
description of the architecture is presented in the supplementary material.

\section{Experimental Results and Discussion}
\label{sec:results}

%The analysis of the results for two steps, the training of the https://www.overleaf.com/projectinformed compression formats which serve as input to train a network for a downstream task. 
We investigate the potential of performing classification,  object detection, few-shot detection and segmentation directly from the different, previously described, informed compressible representations and how well this information generalizes to new tasks and categories. In other words, we investigate whether compression formats can be biased to preserve semantic information. First, we study compression performance and show that our approach is on-par with known state-of-the art CNN-based models and exceeds the performance of popular compression algorithms. 
%while boosting.
%The encoder is frozen and we examine the quality of the learned representation in this regime. This might be included later 
We examine the impact of adding different decoder networks trained to perform computer vision tasks directly from our compressible representation. This includes training directly from compressible codes for semantic segmentation, a held-out task not used to train the representation. We also provide an ablation analysis and examine different model variants. We then explore this approach for learning low shot object detection. Finally, we examine questions of computation efficiency. 
%This effect is further increased when using  representations containing an informed machine perceptual quality.
%Next, we show that ``decoder networks'' trained to perform computer vision tasks directly from an ordinary compressible representation (without semantic regularization) obtain inferior performance compared to training from raw images. This suggests that task-agnostic learned compression schemes do not naturally preserve semantic information needed for tasks like classification, segmentation, and detection.

\subsection{Visual Results and Implementation Details for Learning Compressed Representations}

\begin{figure*}[!ht]
       \centering
         \begin{tabular}{ccc}
         
          \hspace{-.25cm}
                \multirow{2}{*}[4em]{
           \includegraphics[width=0.30\linewidth]{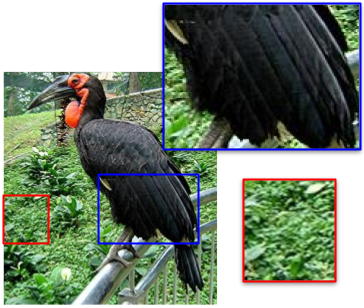}}
            &
          \hspace{-.30cm}
           \includegraphics[width=0.30\linewidth]{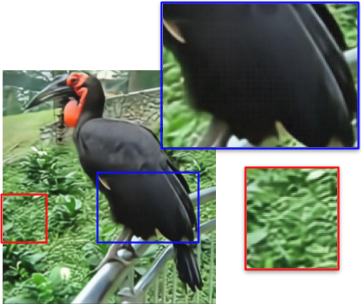} &
           
          \hspace{-.30cm}
          \includegraphics[width=0.30\linewidth]{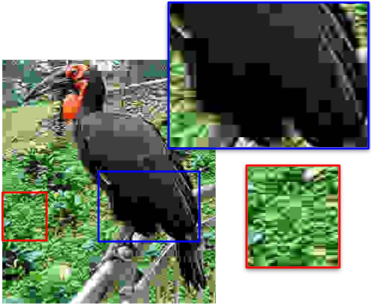} \\
     
          \multirow{2}{*}[-8.3em]{(c) Uncompressed}
          &
          \footnotesize{
          
          \hspace{-.35cm}
          (b) Naive, $D=26.8$, $R = 0.14$} &

          \hspace{-.35cm}
          \footnotesize{(c) JPEG, $D=<26$, $R = 0.31$} * 
          
          \\
          &
          
          \hspace{-.45cm}
          \includegraphics[width=0.30\linewidth]{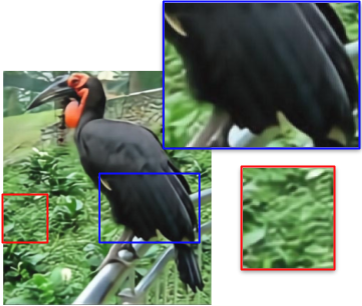} &
          
          \hspace{-.45cm}
          \includegraphics[width=0.30\linewidth]{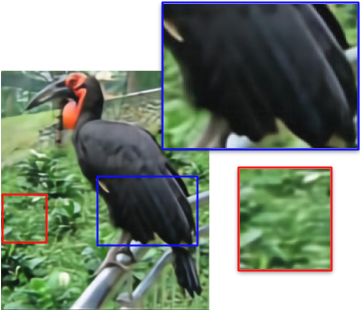}  \\
          &
          
          \hspace{-.75cm}
          \footnotesize{
          (d) $\lambda_t=1k$, $D=26.3$, $R = 0.14$} &
          
          \hspace{-.25cm}
          \footnotesize{(e) $\lambda_t=5k$, $D=25.1$, $R = 0.11$  }

          \end{tabular}
               
       \caption{Visual image quality comparisons, at distortion level $D$ (PSNR), and rate $R$ (bpp). (a) Uncompressed
       ground-truth (b) Naive hyper-prior  based learned compressor. (c) JPEG compressed
       image. (d)-(e) Compressor jointly trained with an object detector using the $\lambda_t$ parameters, respectively, as 1000 and 5000. We observe that increasing the importance of Pascal object detection performance (i.e. the Utility $U$ of the representation) improves the visual quality of the texture features on the birds feathers and reduces the  quality of the background (foliage).}
   \label{fig:detail_quality_compare}
 \end{figure*}

In Figure \ref{fig:detail_quality_compare} we show the results of
our proposed rate-distortion-utility  approach to learned compression in panels (d) and (e). We compare it with JPEG compression at a much higher bitrate of 0.31 bpp, in (c) and a naive version of our approach in (b) -- which one could consider as an instance of, or comparison with the approach proposed in \cite{balle2018variational}, but implemented within a specific encoder architectures held fixed across these experiments. The architecture is the best one from the broader family of neural compression models discussed and explored in more detail below. Notice that we can see a striking phenomenon: at very high $\lambda_t$, the texture on an object category of interest to the classifier/detector (a bird in this case) is preserved while the background (foliage) is compressed. This is in agreement with the recent finding that convolutional networks tend to focus on texture rather than shape \citep{hermann2019origins}. 

%Given equation \ref{eq:semantic_optimization}, we describe in this section, in detail, the process of jointly training compressor given a different set of available tasks.
We obtain a compressible space $z$ to perform vision tasks
by applying Equation (\ref{eq:final_semantic_compression}) over the datasets detailed below. For all our experiments we use the Adam optimizer with an initial learning rate of $5 \times 10^{-5}$. The training dataset used to train the image decoder is a subset of 400k images of OpenImages \cite{kuznetsova2018open} that was pre-processed using the strategy proposed by \cite{mentzer2019practical} that downscales the images in order to reduce potential compression artifacts. To test compression quality results, we use the standard losslessly compressed Kodak dataset \cite{kodak1993kodak}.
We use classification and detection losses to approximate a machine perception quality metric. Differently from \citep{torfason2018towards},
we train on the compression and the machine perception tasks from scratch, jointly. %As a naming convention we refer to compressed representations that were learned jointly with machine perception component as \emph{informed}.
% Classification INformed
For classification, we use multi-class cross-entropy as the loss function. We train the model using the ImageNet dataset jointly with OpenImages \cite{kuznetsova2018open}. We train both tasks for $70$ epochs dividing the learning rate by a factor of 10 every 30 epochs.  In order to sample the same number of images for each task (reconstruction and classification) we
set a batch size of $72$ for image reconstruction, which is substantially
bigger than the batch size used in previous work \cite{balle2018variational}. For the classification task we set a batch size of $256$. Jointly trained with classification we experiment with two \textbf{class informed} models.
We set $\lambda_t=10,\lambda_d=100$ to obtain a bit-rate of $\approx0.4$, $\lambda_t=10,\lambda_d=10$ for a bit-rate of $\approx1.0$.
% Detection Informed
For detection, the task  specific module in this case consist of an architecture based on a Faster-RCNN \citep{ren2015faster} structured module, which contains a region proposal network and bounding box classifier. We use Pascal VOC as the training dataset. Finally, the standard region-proposal regression and detection class loss \citep{ren2015faster} are used as the detection loss function. We built two \textbf{detection informed} models. 
We set $\lambda_t=100,\lambda_d=100$ to obtain a bit-rate of $\approx0.4$ $\lambda_t=100,\lambda_d=25$ for a bit-rate of  $\approx1.0$.
We also trained \textbf{class and detection informed model} with a bit-rate of $\approx1.0$ using  $\lambda_t=5$ for classification $\lambda_t=100$ for detection and $\lambda_d=25$. All the $\lambda$ hyper-parameters were chosen in order to balance the different batch sizes of tasks.

\subsection{Evaluating Our Learned Compression Approach}
We evaluate different strategies for learning informed compressible representations on various downstream tasks. 
For all tasks, the size of the input compressible representation is of $256 \times W/16 \times H/16$. We assume a fully-trained input representation capable being decoded back to the input RGB image; and capability to encode a bitstream from the latent space. In practice the second requirement means that the latent representation is invariant to quantization for the purpose of reconstruction. For all cases, we use the sResnet architecture with a ResNet-101 backbone. We added task specific layers, accordingly, for each task.

%We show the benefits of using an informed representation for three canonical computer vision tasks.
%\subsubsection{Task Specific Training Details}

\textbf{Classification.}
We investigate the classification accuracy on the ImageNet validation set while training in the full ImageNet dataset 2012 \citep{russakovsky2015imagenet}.
We use the default hyper-parameters used for training a Resnet architecture \citep{he2016deep}. This includes, stochastic gradient descent as optimizer and a starting learning rate of 0.1 that is divided by 10 every 30 epochs. We also set the batch size as 256 for training.

\textbf{Detection.}
We also investigate the average precision on a detection task using the Faster-RCNN \citep{ren2016faster} region
proposal and classification heads.
Similarly to segmentation, we initialize
the training with ImageNet weights trained
for the compressed representation. 

\textbf{Semantic Segmentation.}
For this task specific module we use the Deep Lab V3 \cite{chen2017rethinking} and an atrous spatial pyramid pooling layer. 
We train and evaluate a semantic segmentation network on the Pascal VOC dataset \cite{everingham2010pascal}, measuring the mean intersection over union (mIoU) score. It is important to note that competitive semantic segmentation results can only be obtained when starting from backbone weights that were pre-trained on ImageNet, we found this also to be the case when training from compressible representations. Therefore we initialize training for the semantic segmentation task with the sResNet-101 weights learned by the classification task.

%\subsubsection{Results}
%\paragraph{Classification.}
% Explain the downstream task training
%Using hyper-parameter search on the validation set, we found that $\lambda_{t}=10, \lambda_{e}=100$ yield a bit rate of 1 bit/pixel with a distortion rate PSNR of 35.84.
%For all training phases we use Adam as the optimizer and with learning rate of $5 \times 10^{-5}$ and scale the learning rate by a factor of 0.1 every 30 epochs.
% Arquitecture
%As architecture, due to the smaller input size we also use the task decoder architecture to perform the downstream task.
% Other hypers
 
The results when given different representation inputs for training are presented on Table \ref{tbl:representation_v2}. % More details on the training parameters is presented
%on the supplementary material.
% explain some results and conclusions
% When compared to uncompressed
% POINT 1 the best results we see
The main feature we observe is the capability
to improve over the results of an uncompressed
input by using a compressible 
representation that is five times smaller. The uncompressed baselines were trained with the JPEG image set provided on the
ImageNet, for classification, and the PascalVOC datasets, for detection and semantic segmentation. We observe the overall best performance on semantic segmentation and image detection, when using
our compression trained with classification
and detection features. This is an interesting boost and proves the generality of our method
since no features from semantic segmentation
were used when training the compressed representation input. Those results corroborate our hypothesis that constraining the representation
learning problem into also being a compressible
representation aids generalization. Also, it is important to notice that, for segmentation and detection, 
that the classification and detection informed
results perform better than only detection 
or only classification informed compressed formats.
This indicates that increasing the amount of
information for the compressed representation
may improve generalization performance.

% \subsubsection{Training (New) Tasks Straight from Compressible Codes}

\begin{table*}[!ht]

    \resizebox{1\linewidth}{!}{
\footnotesize
\centering
\begin{tabular}{@{}p{3mm}p{59mm}cp{12mm}p{1mm}p{5mm}p{1mm}ccccc@{}}
%\begin{tabular}{@{}l@{\hspace{3mm}}c@{\hspace{3mm}}c@{\hspace{6mm}}c@{\hspace{3mm}}c@{\hspace{9mm}}c@{\hspace{3mm}}c@{\hspace{3mm}}}
 %\begin{tabular}{@{}l@{\hspace{6mm}}r@{\hspace{4mm}}r@{\hspace{4mm}}r@{\hspace{4mm}}r@{}}
\toprule
& 
&& \hspace{-.25cm}Distortion 
& \multicolumn{3}{c}{Segmentation$^*$}  &  \multicolumn{2}{c}{Classification} &&  \multicolumn{2}{c}{Detection}      \\
Rate & Input && PSNR  && mIoU  &&  Top 1   & Top 5 &&  mAP   & AP50 \\   
\midrule %RGB  4.64
& RGB from standard 4 to 5 bpp JPEGs &&   && 70.8        && 77.3  & 93.3 && 53.1 & 80.8   \\
\midrule
\multirow{5}{*}{\rotatebox[origin=c]{90}{bpp $\sim1.0$} } 
& RGB from 1 bpp JPEGs  && \textcolor{red}{32.0}                                && \textcolor{red}{67.5}  && \textcolor{red}{75.0}  & \textcolor{red}{92.2}  && \textcolor{red}{52.3} & \textcolor{red}{79.8} \\
& Our Compression - Naively Learned && 36.2               && {67.5}  && {74.8}  & {92.2}  && {52.8} & {80.3} \\
& Our Compression - Class Informed  && 34.6              && 69.5  && 75.1  & 92.4  && 54.7 & 81.1 \\
& Our Compression - Detection Informed && 35.6          && 70.5  && 74.9  & 92.1  && 54.4 & 81.4 \\
& Our Compression - Class \& Detection Informed  && 34.5  && \textcolor{ForestGreen}{\textbf{71.3}}  && 75.1  & 92.2  &&  \textcolor{ForestGreen}{\textbf{54.6}} &  \textcolor{ForestGreen}{\textbf{81.6}} \\
\midrule
\multirow{4}{*}{\rotatebox[origin=c]{90}{bpp $\sim0.4$} }
& RGB from .4 bpp JPEGs  && \textcolor{red}{28.0}                                   && \textcolor{red}{57.8}  && \textcolor{red}{70.7} & \textcolor{red}{89.8} && \textcolor{red}{51.4} & \textcolor{red}{78.1} \\
& Our Compression - Naive Learned && 30.7               && 68.9  && 74.0 & 91.7 && 52.1 & 79.6 \\
& Our Compression - Class Informed  && 30.1              && 69.0  && 75.1 & 92.3 && 52.8 & 80.1 \\
& Our Compression - Detection Informed  && 30.5          && 70.2  && 74.8 & 92.1 && 52.9 & 80.3 \\
% & Class \& Detection Informed Compression && xx.x  && xx.x & xx.x && xx.x & xx.x \\
\bottomrule
\end{tabular}}
\vspace{3mm}
\caption{Comparison of task performance when using different compressible latent formats as input. Classification experiments are on ImageNet, while Detection and Segmentation experiments are on the Pascal VOC. $^*$Note that none of our learned compressed representations were trained for semantic segmentation; however, a model trained from our best 1bpp compressed representation can even outperform training from high quality RGB images (\textcolor{ForestGreen}{\textbf{bold green}}). Importantly, recall that the encoder is frozen and learning is performed directly from our compressed format.}

\label{tbl:representation_v2}
\end{table*}

% Point 2 our results are even better when  using more compressed spaces and when comparing with the 

We can perceive an even clearer benefit when
training from compressed representations when the compression rate (bits/pixels) is larger. The
image classification results obtained
for low compression, 0.4 bpp, fourth row on Table \ref{tbl:representation_v2}, are able to match
the results obtained for mid 
compression, 1.0 bpp. That indicates a clearer
benefit of using informed compressed spaces in
lower bit-rate regimes.
As a drawback, we see that the improvement
by using informed compressed representations for 
classification is limited, not being able to fully
recover the uncompressed classification results. An insight into 
these results can be acquired by
observing the reconstruction of some informed compressed representations on Fig.\ref{fig:detail_quality_compare} (e) where, even though the bit
rate is lower, the features of the object of interest are well maintained.

\subsection{Ablations and Analysis}
 Table \ref{tab:subpixel_stem} shows the impact that different
 architectural elements have on training
 with compressed representations as input.
 Those differences are compared based on
 the Top 1 classification result on the
 ImageNet validation set. 
We can see that increasing the number of pixel shuffle blocks has a  substantial positive impact on the results. The same is true when adding a residual connection. We also observed some considerable
improvements when using the Mish \citep{misra2019mish}  or the
SiLU \citep{elfwing2018sigmoid} activation functions as compared to the standard ReLU activations.

\begin{table*}[h]
    \centering
    \begin{tabular}{c|c|c|c|c|c}
         One pixel shuffle block & Two pixel shuffle blocks & Residual block & Mish & SiLU & Top 1\\
         \hline
         \checkmark & & & & & 69.2 \\ \hline
         \checkmark & & \checkmark & & & 71.2 \\ \hline
          & \checkmark & \checkmark & & & 74.8 \\ \hline
          & \checkmark & \checkmark & \checkmark& & 75.0\\ \hline
          & \checkmark & \checkmark & & \checkmark & 75.1
    \end{tabular}
    \caption{Ablation analysis and architecture variant comparisons for the classification task.}
    \label{tab:subpixel_stem}
\end{table*}

In Table \ref{tab:vs_joint} we show the comparison our super-resolution based architecture with the one proposed by \cite{torfason2018towards}.
A significant impact on the results can be observed, particularly on the classification task. We believe that this is likely due to a better initial receptive field being presented to the ResNet backbone. We can also see the benefit of using a task informed representation both
for the simpler architecture (cResnet) and the one proposed (sResNet).

\begin{table*}[h]
    \centering
    %\resizebox{1\linewidth}{!}{
    \begin{tabular}{@{}llcccccccc@{}}
    \toprule
    &  && Segmentation  &&  \multicolumn{2}{c}{Classification} &&  \multicolumn{2}{c}{Detection}      \\
    & Input                                          && mIoU  &&  Top 1   & Top 5 &&  mAP   & AP50 \\   
    \midrule %RGB  4.64
    & RGB from standard 4 to 5 bpp JPGs                                     && 70.8  && 77.3  & 93.3 && 53.1 & 80.8   \\
    \midrule
    %& cResnet (0.635 bpp) \citep{torfason2018towards}  && 64.0  && -     & 88.1 && - & - \\
    & cResnet (0.4 bpp) \citep{torfason2018towards}                        && $64.39$       && $70.34$ & $89.75$ && - & - \\    
    & cResnet Detection Informed (0.4 bpp)                 && $64.82$ && $71.57$   & $90.47$ && - & - \\
    % TODO add something like a resnet with upsampling( deconvolution) to be able to cite the other guys
    
    & sResnet (0.4 bpp)                      && 68.9  && 74.0  & 91.7 && 52.1 & 79.6 \\
    & sResnet Detection Informed (0.4 bpp)                      && \textbf{70.2}  && \textbf{75.1}  & \textbf{92.3} && \textbf{52.9} &\textbf{ 80.3} \\
    
    \bottomrule
    \end{tabular}
    % Resizebox }
    %\vspace{3mm}
    \caption{Ablation analysis and comparison of different architectural variants for creating compressed representations on different downstream tasks at low bit-per-pixels.
    } 
    \label{tab:vs_joint}
\end{table*}
\vspace{5mm}

\subsection{New Object Detection Categories in the Low-shot Learning Regime}

The jointly trained compressed representations are particularly interesting for low-shot regimes. We conduct experiments on few-shot detection by reproducing the experiments and evaluation procedure in \citep{wang2020frustratingly} using the their open source code base. The dataset used is Pascal VOC, and Faster-RCNN as the detector architecture \cite{ren2016faster}. The classes are separated into 15 base classes and 5 novel classes. First, the detector is trained on the 15 base classes. Then, in the fine-tuning phase, the detector learns a new bounding-box (bbox) classifier and a new bbox regressor using only $k \in \{1, 2, 3, 5, 10 \}$ samples from the base classes and the novel classes. 
The results of our experiments are summarized on Table \ref{tbl:few_shot_detection_pascal}.
For all the compressed representations we
use a bit rate of 1 bpp.

\begin{table*}[!h]
    %\small
    \centering
    \resizebox{1\linewidth}{!}{
    \begin{tabular}{@{}lcccccccccccccccccc@{}}
        \hline
        \multirow{2}{*}{Method} &&  \multicolumn{5}{c}{Novel Set 1} && \multicolumn{5}{c}{Novel Set 2} &&
        \multicolumn{5}{c}{Novel Set 3}  \\
        $ \quad\quad\quad\quad \# $ of Training Examples   & $\rightarrow$ & 1 & 2 & 3 & 5 & 10
           && 1 & 2 & 3 & 5 & 10
           && 1 & 2 & 3 & 5 & 10 \\
        \midrule
        % YOLO-joint \cite{kang2019fewshot}  
        %      && 0.0 &0.0 &1.8 &1.8 &1.8 && 
        %                 0.0 &0.1 &0.0 &1.8 &0.0 &&
        %                 1.8 &1.8 &1.8 &3.6 &3.9 \\
        MetaDet \cite{wang2019meta} 
             && 18.9 & 20.6 & 30.2 & 36.8 & 49.6 &&
                        21.8 & 23.1 & 27.8 & 31.7 & 43.0 && 
                        20.6 & 23.9 & 29.4 & 43.9 & 44.1 \\
        Meta R-CNN \cite{yan2019meta}  
             && 19.9 & 25.5 & 35.0 & 45.7 & 51.5 && 
                        10.4 & 19.4 & 29.6 & 34.8 & \textbf{45.4} &&
                        14.3 & 18.2 & 27.5 & 41.2 & 48.1\\
        \midrule
        TFA w/fc \citep{wang2020frustratingly}   
            && 36.8 & 29.1 & 43.6 & 55.7 & 57.0 && 
                        18.2 & 29.0 & 33.4 & \textbf{35.5} & 39.0 && 
                        27.7 & 33.6 & 42.5 & 48.7 & \textbf{50.2}\\

        TFA w/cos \citep{wang2020frustratingly} 
            && \textbf{39.8} & 36.1 & 44.7 & 55.7 & 56.0 && 
                        23.5 & 26.9 & 34.1 & 35.1 & 39.1 && 
                        30.8 & 34.8 & 42.8 & 49.5 & 49.8 \\
        \midrule
        TFA w/fc, our compression 
              && 35.6 & 27.4 & 42.8 & 54.2 & 55.6 && 
                        16.3 & 28.0 & 31.9 & 33.9 & 37.8 && 
                        25.0 & 31.5 & 40.6 & 47.0 & 48.4\\

         Ours - Naive 
               && 33.5 & 26.3 & 40.7 & 53.5 & 55.1 && 
                        20.2 & 24.1 & 30.5 & 31.1 & 36.8 && 
                        26.1 & 30.5 & 38.7 & 45.6 & 45.9  \\
        Ours - Class. + Det. Informed  
            && 38.1 & 37.3 & 44.7 & 56.2 & 57.1 && 23.6 & 27.3 & 34.0 & 35.1 & 38.9 && 31.0 & 33.9 & 42.1 & 48.7 & 48.9 \\
        Ours - Classification Informed 
            && 38.8 & 37.2 & 44.1 & 55.3 & 56.7 && 
                        23.7 & 27.2 & 34.5 & 35.3 & 39.2 && 
                        31.2 & 34.9 & 42.5 & 49.4 & 49.5 \\
        Ours - Detection Informed   && 39.3 & \textbf{37.5} & \textbf{44.8} & \textbf{56.3} & \textbf{57.1} && 
                       \textbf{23.7} & \textbf{29.1} & \textbf{34.8} & 35.3 & 39.2 && 
                       \textbf{31.3} & \textbf{34.9} & \textbf{42.9} & \textbf{49.5} & 49.9 \\

    \bottomrule
    \end{tabular}
    }
    \caption{Few-shot detection performance (mAP50) on PASCAL VOC dataset. We compare our results (bottom) when using a informed
    compressed representation with other methods from the literature. We obtain a convincingly better performance while using a mid bit-rate representation (bpp 1.0).}
    \label{tbl:few_shot_detection_pascal}
\end{table*}

 %When lower bpp JPEG images are use to train the original few-shot detector (TFA), the performance drops by approximately 1 mAP50 at 1 bpp and 2 mAP50 at 0.4 bpp. Our compression format without semantic training performs lower then TFA from JPEG using similar compression ratio. 
%
These experiments show that training the few-shot detector from our classification informed compressible representation not only improves overall few-shot performance at 1 bpp, but also achieves state-of-the-art results for many of the experimental configurations. Also, note that the addition of semantic information to the representation has a significant impact, leading to an average increase of 4 points in mAP50. This strongly suggests that, when jointly trained with visual tasks, compressible  representations can serve as a regularizer in low-data regimes. 
Finally, the performance gap between our method
and the TFA method \citep{wang2020frustratingly}
is even larger when the compression
level is increased on the TFA input to match a bit rate of 1.0. 

%We can perceive also, that for low shot regimes,
% This indicates that those trained
% compressible representations are able to hold
% semantic information.

\subsection{Computational and Memory Cost}

All the performance benchmarks were made on 4 Nvidia V100 GPUs processing a batch in parallel.
One of the motivations for our approach is the fact that there exists an advantage to performing downstream tasks directly from a latent compressed space, besides the obvious saving storage/transmission requirement, the input size is considerably smaller and thus computationally cheaper to forward propagate (fprop). 
%In Table \ref{tbl:comp_cost} we compare the number of frames processed per second that the task decoder can fprop given an RGB image input versus our compressed format. 
%
We typically see performance gains of up to $20\%$ on the classification task. However, it should be noted that our experiments are performed at full 32-bit floating point resolution, yet the compressible representation can be expressed using 4-bits. See Appendix I for more details. The use of 4-bit representations for our compression format could make data processing pipelines even more efficient. Furthermore, if combined with appropriately engineered low-bit decoders we suspect that much more dramatic computational gains are possible for those who desire improved performance in this regard. Given that the field of low bit precision deep learning has been making many advances \cite{courbariaux2015binaryconnect,hubara2016binarized,banner2018scalable,iandola2016squeezenet}, low-bit depth decoders based on this type of compression scheme could be an interesting direction for future work. A recent survey in \cite{gholami2021survey} summarizes some key developments, including extremely fast INT8 and INT4 implementations of canonical Convolutional Neural Network architectures. 

% \begin{SCtable}
% %\footnotesize
%     %\centering
%         \caption{Computational performance comparison between the forward pass of an RGB image and a compressed latent input.}
%     \begin{tabular}{@{}lcccc@{}}
%         \toprule
%         Task  && RGB FPS &  Compressed FPS  \\
%         \midrule
%         Classification  && $1716.8$ & $2097.3$ \\ 
%         Semantic Segmentation && $73.7$ & $102.0$ \\
%         Detection && $35.7$ & $39.5$ \\
%         \bottomrule
%     \end{tabular}
%     %\vspace{3mm}

%     \label{tbl:comp_cost}
% %\vspace{-3mm}
% \end{SCtable}

\section{Related Work}
% NEURAL NETWORK BASED COMPRESSION
% Image compression by back propagation: An example of extensional programming 1988
% Image compression with neural networks
%} A survey 1999
% The networks were not trained end to end before, which maybe is a good point.
Early work using auto-encoder architectures for image compression date back to the late 1980's \cite{cottrell1988imagecb} and many extensions were developed during the following decades \cite{jiang1999image}. They follow the traditional approach of pixel transforms, quantization, and entropy coding. The use of recurrent architectures \cite{Toderici2017,johnston2018improved} was proposed to allow configurable rate-distortion trade-off by progressively encoding image residuals, this can also be improved by tiling \cite{minnen2017spatially}.

Most relevant to our work, is the state-of the art technique of creating a latent representation using a variational auto-encoder and using a hyper-prior network to produce a probabilistic model of the quantized latent representation \cite{balle2018variational, minnen2018joint, cheng2020learned}. We use the quantized latent space of the VAE as an input format for visual tasks, thus the latent space should capture all necessary information for performing these tasks. Those techniques
have also been recently extended to include perceptual quality
metrics, drastically improving the
results from a human perception
perspective \cite{mentzer2020high,agustsson2019generative}. The present work does not focus on improving results from a distortion
or human perception viewpoint, but strives to improve the compression format for machine perception via canonical computer vision tasks such as image classification or object detection as proxies.

Recent work has shown that it is possible to learn compressible representations of features in deep neural networks that do not reconstruct input images \cite{singh2020end}, but which serve to perform classifications. Their work shows that it is possible to create image classifiers using the resulting compressed feature representations that have comparable performance to those trained directly on images. 
% COMPRESSED INFERECE
% End-to-end optimized image compression formultiple machine tasks works directly into the image space.
%A number of recent studies have demonstrated that semantic tasks can be performed directly from compressible representations \cite{singh2020end}. 
%
At the other end of the spectrum, as mentioned in our introduction, other work has used the image representations obtained from the ubiquitous JPEG image compression format as input to  convolutional networks. This approach reduces the computational and memory requirements of the overall pipeline of training and classification while achieving a similar accuracy on ImageNet classification \cite{ulicny2017using,fu2016using,gueguen2018faster}. 

We also propose the approach of performing tasks directly from compressible representations; however, the aim of our work is to \emph{learn a semantically sensitive compressible image format} which facilitates downstream tasks such as classification and semantic segmentation.
Of course there is considerable prior work that has involved training auto-encoders to learn image features via low-dimensional embeddings or codes \cite{hinton2006reducing} \cite{masci2011stacked}. However, these types of approaches have not involved explicitly compressing the latent representations and evaluating such methods in terms of compression rates and distortion. In contrast, our approach and architectures produce low dimensional feature spaces with reduced entropy through encouraging representations to be robust to the integer quantization necessary for arithmetic coding. This approach thereby leads to measurable gains in yielding compressible image representations.
% We also have some differences with
% the disentangling literature.  There is a more recent search
%into find disentangling image
%formation criterias %\cite{higgins2017beta}. However we are not specifically interested into the image formation but into a regularized image compressible representation.
% MUlti task learning

Because image reconstruction and other semantic tasks can be viewed as competing objectives, our work is highly related to multi-task learning \cite{ruder2017overview}.
% Our proposed method applies a recently proposed gradient-projection method designed to mitigate interference between tasks \cite{yu2020gradient}. 
% The closest paper to ours
To the best of our knowledge the work most closely related to ours is \cite{torfason2018towards}, which was the first to perform semantic inference directly from a learned, compressible representation. However, in their work performance was compromised on semantic tasks. We also explore inference from a compressible representation, however we aim to develop an algorithm that produces \emph{semantically sensitive compressible representations without sacrificing their ability to compute the information necessary to reconstruct the input image}. Our experiments underscore the various advantages of our proposed format, particularly when learning new out-of-domain tasks and for few-shot generalization. 

\section{Conclusions}
%With the growing prominence of computer vision applications, practitioners should consider compression formats that not only enable high-fidelity reconstructions, but can also be used directly for inference for semantic tasks. 
% We have proposed a simple way of learning a compressed representation that satisfies these requirements.
We have proposed stepping away from the classical rate-distortion paradigm for learned image compression and moving to a rate-distortion-utility framework.
We have shown that this can be achieved by simply jointly training a compressor to also serve as the intermediate representation for multiple tasks. %One of our most interesting findings is that compressed representations trained in this multi-task way provide representations that are particularly effective at low shot learning. 
Our work provides evidence that the combined effects of: (1) compact codes provided by explicit neural compression techniques, and (2) a multi-task learning setup, indeed produces representations that are particularly effective in both: (a) the low shot learning regime, and (b) when training a new task such as semantic segmentation directly from these types of learned compressed representations for machine perception. It should be noted that encouraging compression methods to better preserve certain types of semantic content could introduce biases into which elements of an image are better compressed than others. One must therefore ensure that unwanted biases are not unintentionally induced in the algorithm. If our approach becomes adopted, as time goes on even larger datasets of tasks could address the limitations of training using only the ImageNet, OpenImages and Pascal datasets.

\ifCLASSOPTIONcompsoc
  % The Computer Society usually uses the plural form
  \section*{Acknowledgments}
\else
  % regular IEEE prefers the singular form
  \section*{Acknowledgment}
\fi

We thank NSERC, the COHESA Strategic Network, Google, Mila and CIFAR for their support under the AI Chairs program.

\appendices

\section{Appendix I: Compression Model Details}

\subsection{A Quantized Conditional Gaussian Model}\label{sec:cond_gaussian}
Here we provide some more insight into the probabilistic model used for the arithmetic coding of the quantized compressible representation $\hat{z}$. We view this tensor as simply a list of $N$ elements, where each element is an instance of random variable unique to that element. This means that the probabilistic model over the whole representation can be formalized as 
\begin{equation}
    p_{\hat{z}}(\hat{z}) = \prod\limits_{i=1}^{N}  p_{\hat{z}_i}(\hat{z}_i)
\end{equation} 

Each element $\hat{z}_i$ is modelled by an independent Gaussian with its own mean $\mu_i$ and its own standard deviation $\sigma_i$. Both of these parameters are obtained by decoding $\hat{h}$ using the \textit{hyper-decoder} network as discussed above. %presented in the figure \ref{fig:compression_architecture}. %
We use additive uniform noise to simulate quantization. We can formalize the density function $\hat{z}$ as the following.
\begin{equation}\label{eq:hyper_prior_density}
    p_{\hat{z}_i}(\hat{z}_i) =  F_{\mathcal{N}}\left(\hat{z}_i + \frac{1}{2} \middle| \mu_i, \sigma_i \right) - F_{\mathcal{N}}\left(\hat{z}_i - \frac{1}{2} \middle| \mu_i, \sigma_i \right)
\end{equation}
The cumulative function of a single variable Gaussian is known analytically if both the mean and the variance are known, which is the case here, so it can be used directly.

\begin{figure}[h]
    \centering
    \includegraphics[width=0.45\textwidth]{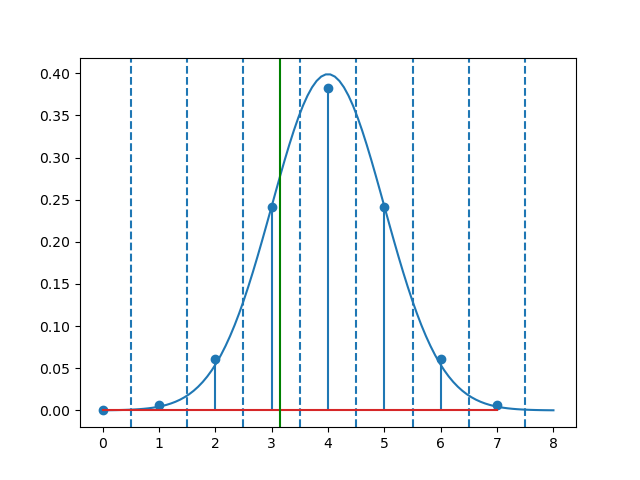}
    \caption{Probability model used for an element of the latent representation if the associated gaussian has a mean of 4 and variance of 1. The key point is that this leads to a latent continuous representation that can be easily quantized into integers in the interval $[0,7]$. }
    \label{fig:gaussaian_quantization}
\end{figure}

To better illustrate the \textit{Conditional Gaussian model}, we take the example of an element of $z$ valued at $3.25$. The corresponding quantized value is 3. To arithmetically encode this value we need a probabilistic model over the integers. The hyper-prior network predicted this number with a Normal distribution with a mean of 4 and a variance of 1. It is not perfect, because the mean could be better and the variance lower. The continuous density function is presented by the blue line in figure \ref{fig:gaussaian_quantization}. To obtain a probability mass function we apply Equation (\ref{eq:hyper_prior_density}) to each of the integers. This is represented by the Diracs in the same image. In practice, we do not track the probabilitic mass of all the integers but only a small subset of of them, specifically the one above a very low hyper-parameter value.  

\subsection{Hierarchical Variational Auto-Encoder}
Now that all the different parts of the compression architecture are defined, we can describe how to actually train them.

\begin{figure*}[h]
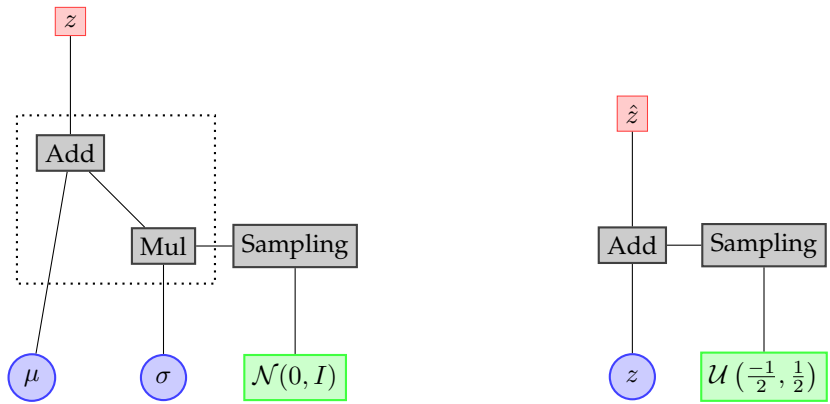

    \centering
    \includestandalone[page=4]{figures/compression/reparam}
    \caption{(left) Computation graph for the sampling procedure used for the reparameterization trick with a Gaussian distribution. (right) Computation graph for the additional use of additive uniform noise to simulate quantization.}
    \label{fig:reparam_uniform}
\end{figure*}

In the univariate case, the added quantization noise around a value can be modelled as uniform density model with the input as the middle point and the quantized representation as a sample from that distribution. We also model each element of the quantized latent representation $(\hat{z},\hat{h})$ as independent, meaning that the density function of the whole density distribution is given by multiplying the density function of all the elements. The resulting density function, given by Equation (\ref{eq:q}), is uniformly distributed in the unit hypercube around the point $(z,h)$ resulting in the processing of a given $x$. 
\begin{equation}
    q(\hat{z}, \hat{h}|x) = \prod_i \mathbbm{1}_{\left[\hat{z}_i -\frac{1}{2}, \hat{z}_i + \frac{1}{2}\right]} \prod_j \mathbbm{1}_{\left[\hat{h}_j -\frac{1}{2}, \hat{h}_j + \frac{1}{2}\right]} 
    \label{eq:q}
\end{equation}
The way to train this density function is to minimize the KL divergence between $q(\hat{z}, \hat{h}|x)$, the learned density function of the quantized representation $(\hat{z},\hat{h})$ and the probabilistic model over it. This is formalized as 
\begin{equation}
    \kld{q}{p_{z,h|x}} = \E_q\left[\log \frac{q(\hat{z},\hat{h}) }{p_(z,h|x)}\right]
\end{equation}
%As seen in annexe \ref{annexe:hyper_prior}, 
This can be reduced to
\begin{equation}\label{eq:hyper_prior_vae_objective}
    \kld{q}{p_{z,h|x}} = -\E_q [\log p(x|\hat{z},\hat{h})] + \kld{q}{p_{\hat{z},\hat{h}}}
\end{equation}
This resulting objective function is the same as the usual VAE objective. To be more precise, it can be seen a Hierarchical Variational Auto-Encoder and $(\hat{z},\hat{h})$, can be seen as a sample of the latent representation of the model. In this view, quantization is equivalent to sampling. This is very similar to the well known ``reparametrization trick'' used for VAEs. The resulting computation graphs for this reparameterization technique are shown in Figure \ref{fig:reparam_uniform}, where $q(\hat{z}, \hat{h}|x)$ is the learned density function of the quantized latent representation, which is an approximation of the true posterior $p(\hat{z}, \hat{h}|x)$.

Equation \ref{eq:hyper_prior_vae_objective} can further be modified to obtain the following form: 
\begin{equation}\label{eq:hyper_prior_loss}
    \kld{q}{p_{z,h|x}} = ||\hat{x}-x||^2 + E_q [- \log p(\hat{z}| \hat{h})] + E_q [- \log p(\hat{h})]
\end{equation}
where $E_q [- \log p(\hat{h})]$ is the expected compression rate using entropy coding of $\hat{h}$ when the univariate density model is used as the  probabilistic model and $E_q [- \log p(\hat{z} | \hat{h})]$ is the expected compression rate using entropy coding of $\hat{z}$ when the Conditional Gaussian model is used as the probabilistic model. 
%Both of these there are cross-entropy losses as presented previously in section \ref{sec:info_theory}.
%
Equation (\ref{eq:hyper_prior_loss}) is not a compression loss because the trade-off between compression and image fidelity cannot be adjusted. So it is modified into the following equation 
\begin{equation}\label{eq:naive_compression}
     \lambda ||\hat{x}-x||^2 + E_q [- \log p(\hat{z}| \hat{h})] + E_q [- \log p(\hat{h})] 
\end{equation}
In this formulation we can easily see the distortion term $||\hat{x}-x||^2$ and the rate terms $E_q [- \log p(\hat{z}| \hat{h})] + E_q [- \log p(\hat{h})]$. This formulation is equivalent to the $\beta$-VAE \cite{higgins2017beta} which has been shown to lead to more disentangled representations, making such representations even more useful as features for other tasks because new tasks could focus on specific features to accomplish specific objectives.

\subsection{Architecture}

\subsubsection{Image Compression framework}
This encoder is composed of four convolutional layers interlaced by Generalized Divisive Normalization (GDN) layers \cite{Balle2016} as activation functions. It was inspired by Divisive Normalization (DN) \cite{carandini2012normalization} which is a multivariate non-linearity designed to model the responses of sensory neurons in biological systems, and it was shown to play a role in efficient information processing. It is defined as:
\begin{equation}
    z_i = \frac{x_i}{(\beta_i + \sum_j \gamma_{ij} |x_j|^{\alpha_{ij}})^{\epsilon_i}}
\end{equation}
where $x_i$ and $z_i$ denote the input and output vectors. The parameters $\alpha_{ij}$, $\beta_i$, $\gamma_{ij}$ and $\epsilon_i$ are trainable parameters. $i$ is the index of the element on which the activation function is computed and $j$ is the index of elements in the neighbourhood of element $x_i$. The neighbourhood is the receptive field of a convolution operation. It  has  been  shown that in practice, GDN, compared to common pointwise activation functions such as relu or tanh,  yields better performance  at  the  same  number  of  hidden  units  and  comes with a negligible increase in number of model parameters when used in an image compression pipeline using neural networks \cite{Balle2016}.
The decoder uses the Inverse Generalized Divisive Normalization (IGDN) and it is defined as :
\begin{equation}
    z_i = x_i (\beta_i + \sum_j \gamma_{ij} |x_j|^{\alpha_{ij}})^{\epsilon_i}
\end{equation}
It uses the same operation as the GDN, but instead if dividing by the normalization factor, it multiplies by it. This is seen as doing the inverse work of the GDN, this is why it is used in the image decoder because it does the inverse function of the encoder. 
We used the same simplifications in the IGDN as the ones we used for the GDN, when it was used in the image decoder.

The encoder expects an RGB image of size $3 \times W \times H$ and outputs a tensor of shape $256 \times W/16\times H/16$. Note, the rank of the compressible format is a third of the input rank and is still much smaller than the input (in bits/pixel) because of its smaller spatial resolution and redundancy. The actual architecture of the encoder and the image decoder are presented in the Table \ref{tab:compression_network_arch}. In this table, \textit{Conv} represents a convolution layer where \textit{k} represents the kernel size, \textit{c} represents the amount of output channels and \textit{s} represent the stride. We use the stride to do the downsampling instead of using a pooling operation. \textit{TConv} is a transposed convolution, it is used to upsample the representation. 

\begin{table}[h]
    \centering
    
    \resizebox{1\linewidth}{!}{
    \begin{tabular}{c|c|c|c}
         Encoder & Image Decoder & Hyper-Encoder & Hyper-decoder  \\
         \hline
         Conv : 5k, 256c, 2s & TConv: 5k, 256c, 2s & Abs & TConv  5k, 256c, 2s \\
         GDN & IGDN & Conv : 3k, 256c, 2s & Leaky ReLU \\
         Conv : 5k, 256c, 2s & TConv: 5k, 256c, 2s & LeakyReLU & TConv  5k, 384c, 2s  \\
         GDN & IGDN & Conv : 5k, 256c, 2s & Leaky ReLU \\
         Conv : 5k, 256c, 2s & TConv: 5k, 256c, 2s & LeakyReLU & TConv  5k, 512c, 2s  \\
         GDN & IGDN & Conv : 5k, 256c, 2s &  \\
         Conv : 5k, 256c, 2s & TConv: 5k, 3c, 2s & &  \\
    \end{tabular}    }
    \vspace{3mm}
    \caption{Architecture of the modules used in the compression framework}

    \label{tab:compression_network_arch}
\end{table}

\subsubsection{Resnet}
The Resnet-101 is used as a baseline backbone for the different vision downstream tasks when going from the RGB images. We present, in detail, on Table \ref{tab:resnets} the truncated-resnet
architecture, used by \cite{torfason2018towards} as compared to the original resnet and our proposed SubPixel variation.

\begin{table*}[h]
    \centering

\includestandalone[]{tables/resnet_now}

    \vspace{3mm}
    \caption{Resnet Architectures. We compare the original Resnet for full scale images (left), with the Truncated Resnet \cite{torfason2018towards} and our SubPixel resnet used on the main results of the paper. }
    \label{tab:resnets}
\end{table*}

\subsubsection{Truncated-Resnet}
When going from the compressible representation, we want to use a backbone architecture that is similar in design and complexity to the Resnet-101 backbone The first architecture that we used is the architecture we call the Truncated-Resnet. 
We cannot use the Resnet stem, which are the first high kernel size convolution blocks on the Resnet architecture. The compressible input is of a different dimensionality and is not compatible with the Resnet backbone. Based on the architecture used in \cite{gueguen2018faster}, we adapt the Resnet backbone by removing the stem, layer 1 and the first two residual blocks of layer 2. The reason we chose to remove those layer is that the spatial resolution of the tensor at this point of the Resnet network matches the spatial resolution of the compressible representation. To match the number of feature maps of the tensor expected at this point in the Resnet architecture, we add a $1 \times 1$ convolutional layer to increase the number of feature maps of the compressible input format. 
We do not change at all the layers 3 and 4 of the Resnet architecture. This way, the number of parameters in the original Resnet architecture and the truncated architecture is very similar, and most of the architecture is the same. This makes it fair to compare performance metrics on the different vision tasks when going from images versus when going from the compressible representation.

\subsubsection{SubPixel-Resnet}\label{sec:subpixel-resnet}
In order to improve on the Truncated-Resnet architecture we develop a subpixel variant. In this new architecture, we keep the layers 1 to 4 the same as in the Resnet architecture, but we propose a new stem that is more adapted to the compressible representation as input. This way, this architecture is mostly the same as the original Resnet architecture, making comparison between them fair.

\begin{figure}[h]
    \centering
    \includegraphics[width=.5\textwidth]{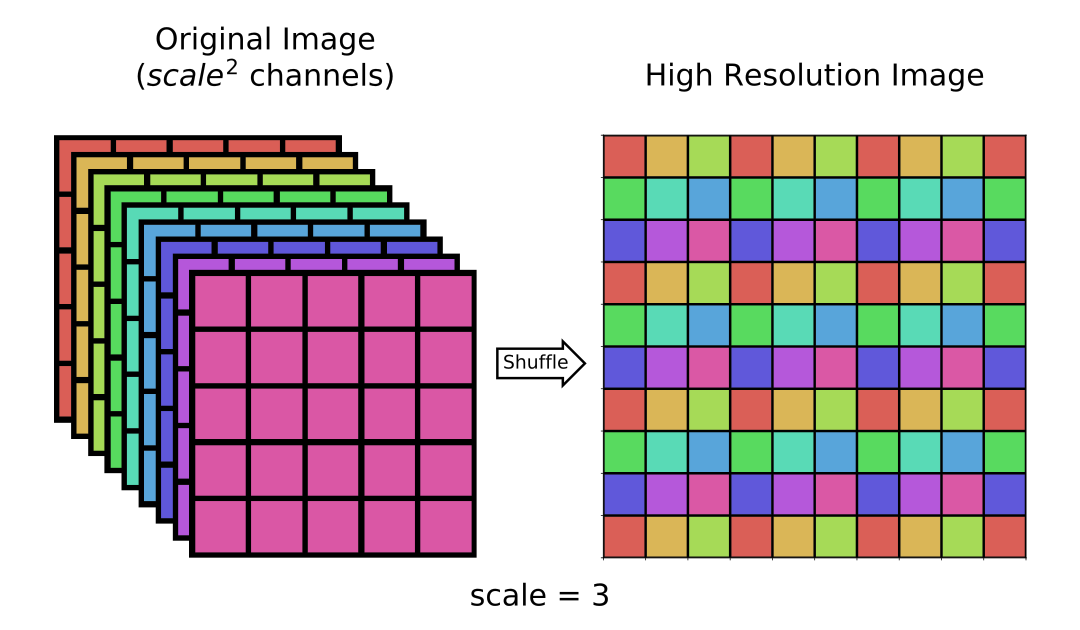}
    \caption{Pixel Shuffle operation with a scale of 3}
    \label{fig:pixel_shuffle}
\end{figure}

First, the expected output tensor spatial dimensionality of the Resnet stem is 4 times smaller than the input image. Our compressible representation has a spatial dimensionality that is 16 times smaller than the input image, which means that we need the output representation of the Sub-Pixel Stem must be four times greater than the ones of the compressible representation. The usual solution in this case is to use a deconvolution operation. We instead chose to use the pixel-shuffle operation followed by convolutions. This operation is recent and was originally proposed in the context of single image super resolution (SISR). The goal of that research was to produce a high-resolution image from a low-resolution input and they achieved state-of-the-art results at the time \cite{shi2016realtime}. The process is shown in figure \ref{fig:pixel_shuffle} and is called a sub-pixel convolution when it is followed by a convolution operation. 
 The way it works is by moving the low-resolution channels into the spatial dimension to create an image with a higher spatial resolution. This is a more efficient transformation then the transposed convolution operation. It does not increase the receptive field, this is why it called a sub-pixel convolution. Also, it has been shown \cite{chen2017rethinking}, that the sub-pixel convolution has a better than the deconvolution. The deconvolution creates checkerboard artifacts in a multitude of tasks when it is used to increase the spatial resolution. To alleviate this checkerboard problem a solution that was used what to upsample the representation using bi-cubic interpolation and then do a convolution. The sub-pixel convolution does not create checkerboard artifacts and replaces the up-sampling steps. We separate the 4 times spatial up-sampling in two separate operations that will each up-sample by a factor of two. This is done, because preliminary results showed that it was yielding better performance.

Second, we change the convolution in the original stem with a custom residual block, similar to the one used in Resnet architecture, but only composed of convolutional layers with a kernel of size 3. %It is presented in the figure \ref{fig:sub_pixel_block}.
The three consecutive convolutions with kernel size of 3 have the same receptive field as a single convolution layer with a kernel of 7. This replaces the convolution in the stem of the original Resnet architecture, while having fewer parameters. This type of replacement has been shown to improve performance in vision tasks both in performance metrics and in speed. The normalization function we used is batch normalization.

% biography section
% 
% If you have an EPS/PDF photo (graphicx package needed) extra braces are
% needed around the contents of the optional argument to biography to prevent
% the LaTeX parser from getting confused when it sees the complicated
% \includegraphics command within an optional argument. (You could create
% your own custom macro containing the \includegraphics command to make things
% simpler here.)
%\begin{IEEEbiography}[{\includegraphics[width=1in,height=1.25in,clip,keepaspectratio]{mshell}}]{Michael Shell}
% or if you just want to reserve a space for a photo:

%\begin{IEEEbiography}{Michael Shell}
%Biography text here.
%\end{IEEEbiography}

% if you will not have a photo at all:
%\begin{IEEEbiographynophoto}{John Doe}
%Biography text here.
%\end{IEEEbiographynophoto}

% insert where needed to balance the two columns on the last page with
% biographies
%\newpage

%\begin{IEEEbiographynophoto}{Jane Doe}
%Biography text here.
%\end{IEEEbiographynophoto}

% You can push biographies down or up by placing
% a \vfill before or after them. The appropriate
% use of \vfill depends on what kind of text is
% on the last page and whether or not the columns
% are being equalized.

%\vfill

% Can be used to pull up biographies so that the bottom of the last one
% is flush with the other column.
%\enlargethispage{-5in}

% that's all folks
\end{document}

%% file: figures/compression/reparam.tex
\tikzstyle{leaf}=[circle,thick,draw=blue!75,fill=blue!20,minimum size=6mm]
\tikzstyle{fixed_leaf}=[thick, draw=green!75, fill=green!20, minimum size=6mm]
%\tikzstyle{root}=[place,draw=red!75,fill=red!20]
\tikzstyle{root}=[draw=red!75,fill=red!20]
\tikzstyle{post}=[]
\tikzstyle{operation}=[rectangle,thick,draw=black!75, fill=black!20,minimum size=4mm]

% gaussian
\begin{tikzpicture}[node distance=1.75cm,>=stealth,bend angle=45,auto]
    \node [root] (sample)                        {$z$};
    \node [operation] (add) [below of=sample]    {Add}
        edge [post]                  (sample);
    \node [operation] (mul) [below right of=add]      {Mul}
        edge [post]                  (add);
    \node [operation] (sampling) [right of=mul]      {Sampling}
        edge [post]                  (mul);
    \node [fixed_leaf] (sampling_param) [below of=sampling] {$\mathcal{N}(0,I)$}
      edge [post]                  (sampling);
        
    \node [leaf] (sigma) [below of=mul] {$\sigma$}
      edge [post]                  (mul);
      
    \node [leaf] (x) [left of=sigma] {$\mu$}
      edge [post]                  (add);
    \usetikzlibrary{calc}
    \draw[black,thick,dotted] ($(add.north west)+(-0.25,0.25)$)  rectangle ($(mul.south east)+(0.25,-0.25)$);
\end{tikzpicture}

\hspace{3cm}

% uniform
\begin{tikzpicture}[node distance=1.75cm,>=stealth,bend angle=45,auto]
    \node [root] (sample)                        {$\hat{z}$};
    \node [operation] (add) [below of=sample]    {Add}
        edge [post]                  (sample);
    \node [operation] (sampling) [right of=add]      {Sampling}
        edge [post]                  (add);
    \node [fixed_leaf] (sampling_param) [below of=sampling] {$\mathcal{U}\left(\frac{-1}{2},\frac{1}{2}\right)$}
      edge [post]                  (sampling);
    \node [leaf] (x) [below of=add] {$z$}
      edge [post]                  (add);
\end{tikzpicture}

%% file: tables/resnet_now.tex
\newcommand{\stem}[0]{
    \begin{bmatrix}
    7 \times 7, 64  \text{, stride 2}\\
    3\times 3 \text{, MaxPool, stride 2}
    \end{bmatrix}
}

\newcommand{\resA}[0]{
    \begin{bmatrix}
    1 \times 1, 64\\
    3 \times 3, 64 \\
    1 \times 1, 256
    \end{bmatrix}
}
\newcommand{\resB}[0]{
    \begin{bmatrix}
    1 \times 1, 128\\
    3 \times 3, 128 \\
    1 \times 1, 512
    \end{bmatrix}
}
\newcommand{\resC}[0]{
    \begin{bmatrix}
    1 \times 1, 256\\
    3 \times 3, 256 \\
    1 \times 1, 1024
    \end{bmatrix}
}
\newcommand{\resD}[0]{
   \begin{bmatrix}
    1 \times 1, 512\\
    3 \times 3, 512 \\
    1 \times 1, 2048
    \end{bmatrix}
}

% layers shapes
\newcommand{\layerOneShape}[0]{  $64 \times 64 \times 64$ }
\newcommand{\layerTwoShape}[0]{  $256 \times 64 \times 64$ }
\newcommand{\layerThreeShape}[0]{$512 \times 32 \times 32$ }
\newcommand{\layerFourShape}[0]{ $1024 \times 16 \times 16$ }
\newcommand{\outputShape}[0]{ $2048 \times 8 \times 8$ }

% layers Arch
\newcommand{\layerOneArch}[0]{  $\resA \times 3$ }
\newcommand{\layerTwoArch}[0]{  $\resB \times 4$ }
\newcommand{\layerThreeArch}[0]{$\resC \times 23$ }
\newcommand{\layerFourArch}[0]{ $\resD \times 3$ }

\newcommand{\layerTwoArchTruncated}[0]{
   $\begin{bmatrix}
    1 \times 1, 512 \\
     \resB \times 2 
    \end{bmatrix}$
}

\newcommand{\compressedStemArch}[0]{
   \begin{bmatrix}
    \text{Block(256, 64)}\\
    \text{PixelShuffle(2)} \\
    \text{Block(64, 16)}\\
    \text{PixelShuffle(2)} \\
    \text{Block(16, 16)}\\
    1 \times 1, 64
    \end{bmatrix}
}
\hspace{-10mm}

\begin{tabular}{c|c|c|c}

Block name  & Resnet          & Truncated Resnet       & SubPixel Resnet       \\
\hline
Stem        & $\stem$         & -                      & $\compressedStemArch$ \\
\hline
layer2      & \layerOneArch   & -                      &  \layerOneArch        \\
\hline
layer3      & \layerTwoArch   & \layerTwoArchTruncated & \layerTwoArch         \\
\hline
layer4      & \layerThreeArch & \layerThreeArch        &  \layerThreeArch      \\
\hline
layer5      & \layerFourArch  & \layerFourArch         & \layerFourArch        \\
\end{tabular}